\documentclass{article}

\usepackage{hyperref}
\usepackage{arxiv}
\usepackage[utf8]{inputenc} 
\usepackage[T1]{fontenc}    
\usepackage{hyperref}       
\usepackage{url}            
\usepackage{float}
\usepackage{algorithm}
\usepackage{algpseudocode}
\usepackage{amsmath}
\DeclareUnicodeCharacter{22EF}{\dots}

\usepackage{booktabs}       

\usepackage{amsfonts}       
\usepackage{nicefrac}       
\usepackage{microtype}      
\usepackage{lipsum}
\usepackage{graphicx}
\graphicspath{ {./images/} }
\usepackage{amssymb} 
\usepackage{multicol}

\title{Antimagic Labeling of Graphs Using Prime Numbers}

\author{
 Arafat Islam \\
  Department of Computer Science\\
  American International University-Bangladesh (AIUB)\\
  \texttt{19-39377-1@student.aiub.edu} \\
   \And
 Md. Imtiaz Habib \\
  Department of Computer Science\\
  American International University-Bangladesh (AIUB)\\
  \texttt{19-39389-1@student.aiub.edu} \\
}

\begin{document}
\maketitle
\begin{abstract}
Graph labeling is a technique that assigns unique labels or weights to the vertices or edges of a graph, often used to analyze and solve various graph-related problems. There are few methods with certain limitations conducted by researchers previously on this topic. This research paper focuses on antimagic labeling of different types of graphs and trees. It entails the assignment of distinct prime values to edges in a manner that ensures the cumulative sum of edge labels at each vertex remains unique. This research proposes a conjecture on antimagic labeling of any graphs and proves two theories. Firstly, we tried to give weights to the edges randomly, as some exceptions are faced in particular phases in this way, we followed a whole new way to mitigate this problem. This research paper demonstrates computational and mathematical verification to prove that antimagic labeling of any perfect binary tree and complete graph is possible. 
\end{abstract}
\hspace{3.5em}\textit{\textbf{Keywords}: Graph; Antimagic; Graph Labeling; Binary Tree; Complete Graph; Prime Number}

\noindent\makebox[\linewidth]{\rule{\textwidth}{0.2pt}}
\setlength{\oddsidemargin}{-0.5in} 
\setlength{\evensidemargin}{-0.5in} 
\setlength{\columnsep}{1cm}
\begin{multicols}{2}
\section{Introduction}
In recent years, graph labeling has become a prominent research topic in the field of graph theory and algorithm. Previously, many researches have been conducted over different kind of labeling for different kind of graph. 

In \cite{r1}, researchers have demonstrated that antimagic labeling is possible for canonical decomposition graphs. Nevertheless, very few researches have been conducted over antimagic labeling for graphs and no research is not found where there is no exception. Hartsfield and Ringel's research \cite{r2} demonstrated the antimagic nature of graph types $P_{n}$ (path graphs), $S_{n}$ (star graphs), $C_{n}$ (cycle graphs), $K_{m}$ (complete graphs), $W_{m}$ (wheel graphs), and $K_{2}$, $m$ (complete graphs with $m\geqslant3$). Additionally, they proposed two conjectures concerning
antimagic labeling of graphs which are still open. It says,Every connected graph except graphs with two vertices has an antimagic labeling. Also, every tree except trees with two vertices has an antimagic labeling \cite{r2}.

According to \cite{r3}, antimagic labeling for any perfect binary tree is possible except for some specific situations. This research paper focuses to demonstrate the proof both mathematically and practically of antimagic labeling of any perfect binary and provide possibility of antimagic labeling on other graphs based upon certain criteria. This research paper aims to provide comprehensive understanding of antimagic labeling of various graph types, focusing on the use of prime numbers and the importance of pattern-based labeling. By building our understanding of these labeling techniques, we contribute to the expanding landscape of graph theory and offer potential directions for further exploration in algorithm and graph theory. 

In this research, we have proposed a conjecture regarding antimagic labeling for almost every graph following an order using prime numbers.

\section{Background} \label{Background}
A graph is a non-linear data structure consisting of vertices and edges. A graph, denoted as $G = (V, E)$, also has endpoints that connect two vertices with each edge \cite{r4}. In other words, a graph can be represented using two finite sets, $V$ and $E$, where $V$ is referred to as the set of vertices, and $E$ as the set of edges. Graphs are categorized into various types based on their properties, such as Trees, Complete Graphs, Cycle Graphs, Path Graphs, Bipartite Graphs, among others \cite{r5}.

A graph, denoted as $G$, can be specifically classified as a Bipartite Graph when its vertices can be divided into two separate and independent sets, labeled as $U$ and $V$. The defining feature of a Bipartite Graph is that each edge $(u, v)$ connects either a vertex from set $U$ to a vertex from set $V$ or vice versa, linking a vertex from set $V$ to a vertex in set $U$. To clarify the concept further, as mentioned in \cite{r6}, a graph $G_{u1,u2}$ is also considered bipartite if its vertex set can be partitioned into two disjoint sets, $U_{1}$ and $U_{2}$, in such a way that for any pair of vertices $(u_{1i}, u_{1k} \in U_{1})$ and $(u_{2i}, u_{2k} \in U_{2})$, no edge $e \in E$ exists where $e$ is of the form $(u_{1i}, u_{1k})$ or $(u_{2i}, u_{2k})$. Moreover, it has been portrayed in \cite{r7} that a graph $G$ can exclusively be classified as bipartite if and only if it contains no odd cycles as subgraphs. This condition serves as a decisive criterion for determining the bipartite nature of a given graph. In simpler terms, for every edge $(u, v)$, either $u$ belongs to set $U$ and $v$ to set $V$, or $u$ belongs to set $V$ and $v$ to set $U$.

A tree is a connected graph without cycles. Particularly, the tree is a highly researched and practically used class of graphs that has gathered significant attention in numerous studies \cite{r8}. A connected acyclic graph is considered as a tree \cite{r4}. Also, a forest is a graph whose components are all trees \cite{r7}. Every two vertices of a tree are connected by exactly one unique path. In other words, a graph is a tree if and only if it does not contain any loop and precisely has single unique path. 

Binary trees are distinguished by a defining attribute where every node has precisely two children: a left child and a right child \cite{r9}. In other words, a tree is called a binary tree if the parent nodes of the tree contain no more than two children. The children are denoted as left child and right child based upon their positions \cite{r9}. In binary trees, a node is considered as leaf if it does not carry any child node. Binary trees can be classified into various types, such as, Balanced Binary Tree, Full Binary Tree, Complete Binary Tree and Perfect Binary Tree. A binary tree is balanced if the difference of height between left sub-tree and right sub-tree is either 0 or 1. Both left and right sub-tree of that node is balanced. A binary tree is said to be a complete binary tree when all the levels of the tree are filled except for deepest level nodes which are apparently leaves and filled accordingly from left to right \cite{r10}. Another type of tree is full binary tree, where it is not mandatory for a tree to have leaves at the lowest level. A full binary tree exactly has 0 or 2 child nodes. Lastly, a perfect binary tree represents a special type of binary tree where all the nodes of the tree are filled with exactly 2 child nodes except for the leaves which are at the same deepest and same level of other leaves in that tree. A perfect binary tree has cousins that share a common grandfather and ancestor only if the father is not the root node. Also, a perfect binary tree contains at least one brother. 

Antimagic labelling is a type of labeling scheme used in graph theory. A graph $G$ is considered as anti-magic if it consists an antimagic labeling \cite{r2}. A graph is considered planar if it possesses the property of being amenable to representation within the two-dimensional plane in such a manner that no two edges intersect in a nontrivial manner, except at a common vertex to which they are both directly connected \cite{r11}.

\section{Related Works}  \label{Related-Works}
Hartsfield and Ringel introduced the concept of antimagic labeling for graphs. They demonstrated that paths, 2-regular graphs, and complete graphs possess anti-magic properties, and they also proposed two concepts related to the labeling of graphs with anti-magic properties \cite{r2}. It involves assigning unique numbers or labels to the edges or vertices of a graph in a way that ensures that the sums of the labels on the edges incident to each vertex are all different. In other words, the sum of the labels on the edges connected to any vertex in the graph will be a distinct number. This unique property makes the graph antimagic, and it has been studied by researchers in graph theory due to its interesting properties and applications. 

Previously, few notable researches have explored diverse methods of implementing antimagic labeling on various types of graphs. Nevertheless, most of them have some limitations and exceptions. For instance, in \cite{r3}, positive real numbers were used orderly to label the edges in order satisfy the conditions of antimagic labeling. But there was one exception mentioned in \cite{r3} that antimagic labeling is possible except for the levels that follow 3,7,11…. series. After applying a swap operation between those edges with particular edges, they were able to satisfy the conditions to maintain antimagic labeling. An example of this swap is given below. 
\begin{figure}[H] 
    \centering
    \includegraphics[height=2.3cm]{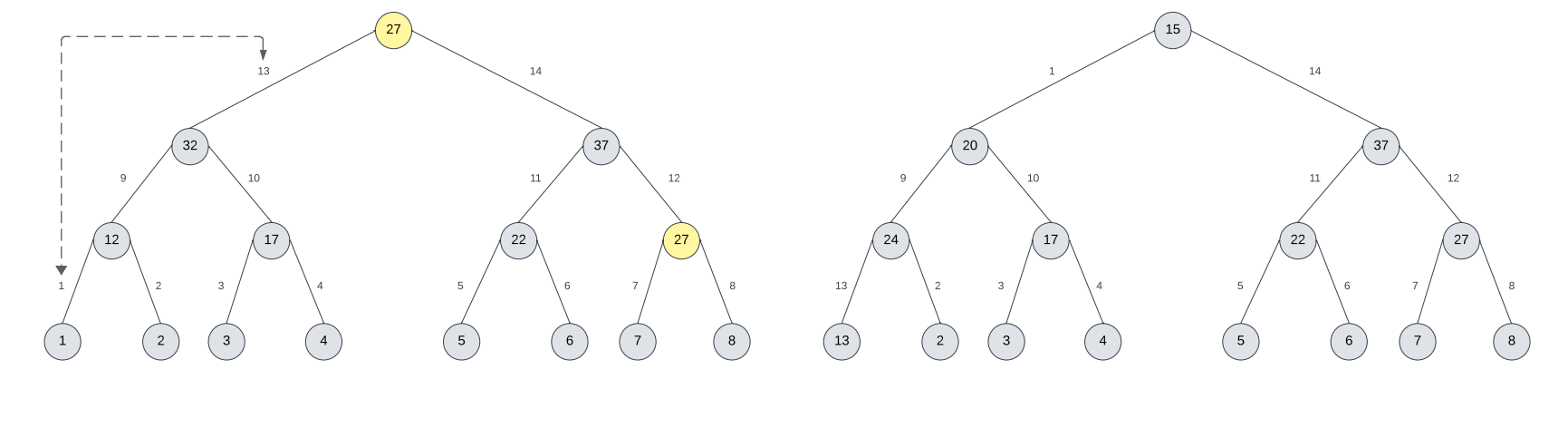}
    \caption{Swap Edges Labels of Perfect Binary Tree}
    \label{fig:1}
\end{figure} 
We have proved by implementing antimagic labelling of any binary tree using only prime numbers without exception. A prime number is defined as a positive integer that is irreducible to every natural number which is less than itself. In simpler terms, a prime number is a number that cannot be divided by any smaller natural number except for 1 and itself. This definition is different from the traditional definition of prime numbers based on divisibility \cite{r12}. 
Another example would be cubic graphs, which are graphs where every vertex has a degree of three, signifying that each vertex is connected to precisely three edges. In \cite{r13}, the authors have demonstrated the antimagic nature of cubic graphs. They have established a direct relationship between the edges of a cubic graph and the integers ranging from 1 to the total number of edges. This relationship ensures that the sums of labels assigned to edges connected to different vertices are distinct. However, it's important to note that their work is specifically focused on cubic graphs and does not address the potential for antimagic labeling in graphs other than the cubic graphs.

In graph theory a regular bipartite graph is a graph in which the vertices can be separated into two separate sets, and every edge in the graph links a vertex from one set to a vertex in the other set. According to \cite{r14}, Regular bipartite graphs are characterized by having a balanced structure, with an equal number of vertices in each set and an equal number of edges connecting the two sets. The authors demonstrated that every regular bipartite graph with a degree of at least 2 is antimagic, using the Marriage Theorem as a key technique. Although it does not mention about the possibility of antimagic labeling on different types of graphs based upon their methodology, it indicates that different types of regular graphs may require different approaches in their analysis and it explicitly mentions that the proof for 4-regular graphs is more complicated than for 6-regular graphs \cite{r14}. In addition to that, the paper does not provide a proof for the antimagic property of regular bipartite graphs with degrees larger than 2.
Also, A regular graph with an even degree that is not bipartite is a specific kind of graph. This regular graph has all its vertices with the same degree, and it is non-bipartite, implying that it cannot be divided into two separate sets of vertices in a way that all edges connect vertices from different sets \cite{r15}.  

According to \cite{r16}, A generalized pyramid graph is formed by connecting every vertex of a base graph to a central vertex known as the apex. The base graph can be any graph with a consistent degree of k and p vertices. This resulting graph is represented as P(G, m), with G referring to the base graph and m indicating the pyramid's level. 
Research conducted by Alon, Kaplan, Lev, Roditty, and Yuster, initially proved that all the complete partite graphs (excluding K2) and graphs that contain a maximum degree of at least (n-2) possess antimagic properties \cite{r17}. Later, the authors fixed the error in the initial proof, affirming the validity of the anti-magic labeling theorem for trees that include at most one vertex with a degree of 2 in \cite{r18}.

Apart from that, researches have been conducted over antimagic labeling on are various types of graphs. Such as, the authors of \cite{r19} determine the local antimagic total chromatic number, for amalgamation graphs of complete graphs. Similarly, in \cite{r20} the existence of distance antimagic labeling in the ladder graph is investigated.

\section{Conjecture} \label{conjecture}

Inspired  from the characteristics of prime numbers, we have proposed a
conjecture about the possibilities of antimagic labeling. \\
\\ \textbf{Conjecture 1:} Every graph is antimagic when it follows an
order of edge labeling using prime numbers except graphs with less than 3
nodes.\\

In the following sections, the proof of the above conjecture for different
graphs have been discussed.
\section{Perfect Binary Tree} \label{perfect-binary}
To prove antimagic labeling is possible for a perfect binary tree, we have approached
with a distinct way. Using prime numbers as weights of the edges of a perfect binary
tree, satisfies all the conditions that antimagic labeling has. First, we implemented
a function to generate prime numbers to label the edges. In this research paper we
have proved that antimagic labeling is possible for any perfect binary tree using
prime numbers without any exception both mathematically and implemented the algorithm
using python. 

In this section, we have discussed about algorithm. 
We have initiated our work by traversing the perfect binary in a bottom-to-top
manner. In this way, the leaves of the tree come first to get labeled then all the
way to root. This traversal method is one of the most crucial part of our approach.
After that, Left-to-right horizontal strategy is used to label the perfect binary
tree. This strategy completely aligns with our goal and guarantees the uniqueness of
labels. 
To generate prime numbers, we have applied two distinct algorithms that ensure the
prime numbers used to label the edges are unique and suitable. Furthermore, another
algorithm is used to label the edges in left-to-right, bottom-to-top approach which
also ensures that weights of all the edges in the perfect binary tree are
distinctive. Next, the mathematical verification will be analyzed with algorithms. \\

\begin{figure}[H] 
    \centering
    \includegraphics[height=3.5cm]{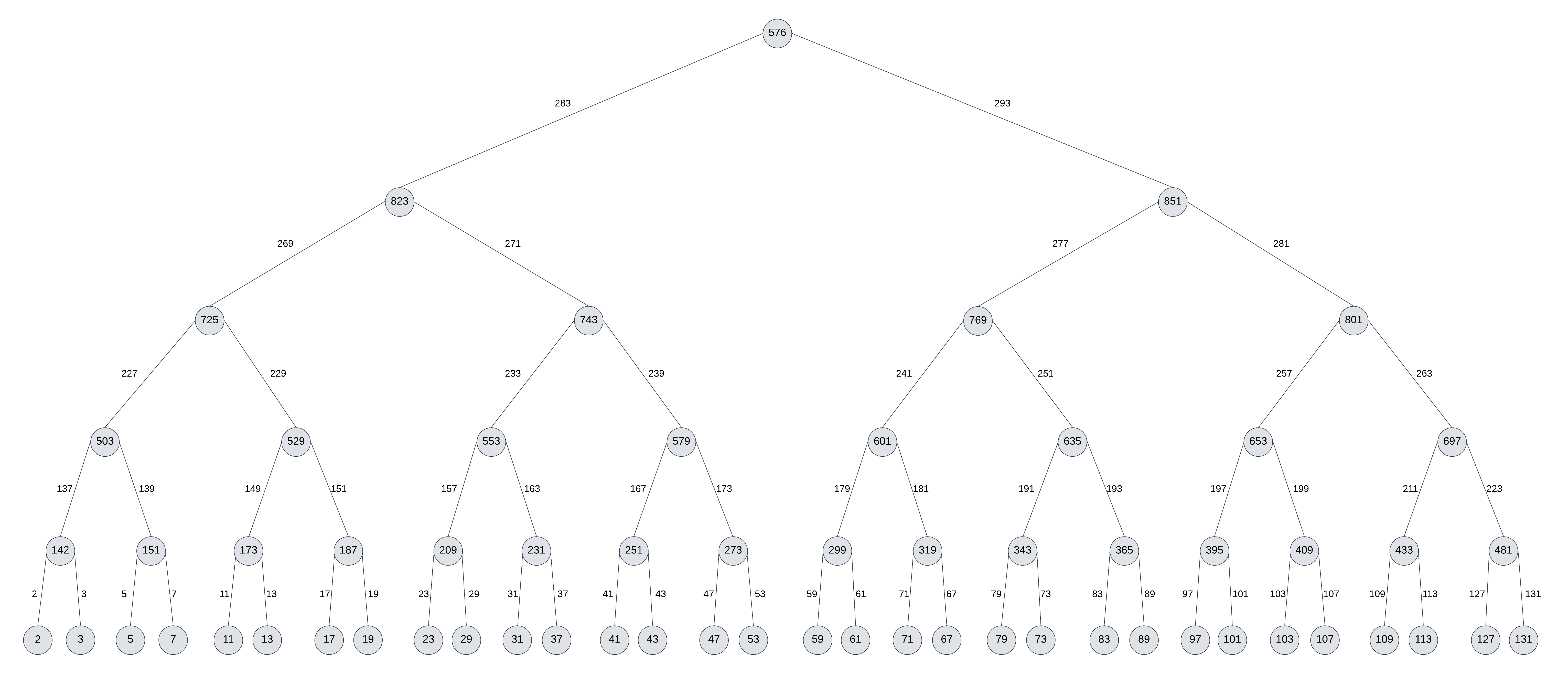}
    \caption{Antimagic labeling for perfect binary tree of level 5}
    \label{fig:6}
\end{figure} 

\subsection{Pseudocode:}
\subsubsection{Algorithm 1: Pseudocode for generating prime numbers}
\begin{algorithm} [H]
\caption{Generate Primes}
\begin{algorithmic}[1]
\Procedure{GeneratePrimes}{$\text{limit}$}
\State $\text{primes} \gets [\:]$
\State $\text{is\_prime} \gets \text{True} \times (\text{limit} + 1)$
\State $\text{is\_prime}[0] \gets \text{False}$
\State $\text{is\_prime}[1] \gets \text{False}$
\State $\text{num} \gets 2$
\While{$\text{length of primes} < \text{no\_of\_edges}$}
    \If{$\text{is\_prime[num]}$ is True}
        \State $\text{append num to primes}$
        \For{$\text{multiple}$ in $\text{range}(\text{num}^2, \text{limit} + 1, \text{num})$}
            \State $\text{is\_prime[multiple]} \gets \text{False}$
        \EndFor
    \EndIf
    \State $\text{num} \gets \text{num} + 1$
\EndWhile
\State \Return $\text{primes}$
\EndProcedure
\end{algorithmic}
\end{algorithm}

\subsubsection{Algorithm 2: Pseudocode for generating prime numbers}

\begin{algorithm} [H]
\caption{Generate Primes}
\begin{algorithmic}[1]
\Function{GeneratePrimes}{$\text{level}$}
\Function{IsPrime}{$n$}
\If{$n < 2$}
    \State \Return false
\EndIf
\For{$i$ from $2$ to $\sqrt{n}$ (inclusive)}
    \If{$n \bmod i = 0$}
        \State \Return false
    \EndIf
\EndFor
\State \Return true
\EndFunction
\State $\text{primes} \gets []$
\State Initialize $n$ to $0$
\State Set $\text{target\_count} = (2^{\text{level}} \times 2) - 2$
\While{length of $\text{primes} < \text{target\_count}$}
    \If{IsPrime($n$)}
        \State Append $n$ to $\text{primes}$
    \EndIf
    \State Increment $n$ by $1$
\EndWhile
\State \Return $\text{primes}$
\EndFunction
\end{algorithmic}
\end{algorithm}

\subsubsection{Algorithm 3: Pseudocode for generating perfect binary trees}
\begin{algorithm}[H]
\caption{PerfectBinaryTree}
\begin{algorithmic}[1]
\Procedure{PerfectBinaryTree}{$\text{level}$}
\If{$\text{level} \leq 0$}
    \State \Return 0
\EndIf
\State $\text{edges} \gets \text{generateprimes}(\text{level})$
\State $\text{new\_edges}, \text{latest\_lower\_edges}, \text{latest\_nodes} \gets \text{lastlevelnode}(\text{edges}, \text{level})$
\State $\text{all\_nodes}.\text{extend}(\text{latest\_lower\_edges})$
\State $p \gets \text{level} - 1$
\While{$p \geq 0$}
    \If{$p > 0$}
        \State $\text{new\_edges}, \text{latest\_upper\_edges} \gets \text{generateedge}(\text{new\_edges}, p)$
\State $\text{nodes} \gets \text{generatenode}(\text{latest\_upper\_edges},$ \newline
\hspace{4em} $\text{latest\_lower\_edges})$
        \State $\text{latest\_lower\_edges} \gets \text{latest\_upper\_edges}$
        \State $p \gets p - 1$
    \EndIf
    \If{$p = 0$}
        \State $\text{root\_node} \gets \text{getrootnode}(\text{new\_edges})$
        \State \textbf{break}
    \EndIf
\EndWhile
\EndProcedure
\end{algorithmic}
\end{algorithm}

\subsection{Mathematical Verification:}
In this section, we present a mathematical demonstration of the feasibility of an anti-magic labeling scheme for perfect binary trees, using prime numbers as labels. 

The essence of our approach lies in the systematic labeling of all edges within the tree. By following a sequential left-to-right and bottom-to-up order, we assign unique prime number labels ranging from $1 to e_l$, where l represents the level of the tree, and el signifies the total count of edges at that specific level.

Throughout the research we have created four distinct formulas. These formulas enable us to determine the value and position of vertices and edges within a perfect binary tree relative to its level. Our established methodology not only underscores the plausibility of anti-magic labeling but also underpins the logical foundation upon which our proof is constructed. Through a comprehensive analysis of these formulas, we substantiate the viability of our proposed anti-magic labeling approach.\\

Number of potential vertices within a perfect binary tree of level,
\begin{equation}
    v_l=2^l+2^{l-1}+2^{l-2}+⋯+2^0  \label{eq:1}
\end{equation} \\

Number of potential edges within a perfect binary tree of level,
\begin{equation}
    e_l=(2^l*2)-2 \label{eq:2}
\end{equation} 

To establish the feasibility of antimagic labeling for all perfect binary trees, our initial step involves categorizing such trees into four distinct groups based on their vertex characteristics.

\begin{itemize}
\item Root Node: \\
\begin{equation}
    r(l)= P_1[(2^l*2)-3]^{th} + P_2[(2^l*2)-2]^{th} \label{eq:3}
\end{equation}

\item Second-To-Last:
\begin{equation}
    n(l,k)=P_1 (2n-1)^{th} + P_2(2n)^{th} + P_3[(2^l)+n]^{th} \label{eq:4}
\end{equation}

\item Last Level:
\begin{equation}
    P(n)=1\geq n \leq 2^l \label{eq:5}
\end{equation}

\item Internal Levels: (when $k < 1$)
\begin{equation}
\begin{split}
    n(l,k) = & P_1\left[(2n-1) + \sum_{i=0}^{k-2} 2^{l-i}\right]^{th} + \\
    & P_2\left[(2n) + \sum_{i=0}^{k-2} 2^{l-i}\right]^{th} + \\
    & P_3\left[\sum_{i=0}^{k-1} 2^{l-i} + n\right]^{th}
\end{split}
\label{eq:6}
\end{equation}
\end{itemize} 
Where, \\
${l}$= Number of levels of the tree. \\
${n}$= Position of the node within a level (Left-To-Right).\\
${k}$= Targeted level within the tree (Bottom-To-Top). \\
${n(l,k)}$= This represents the value of the node located in the level K of the L-level perfect binary tree at position n, considering the edge position $P_1, P_2$ and $P_3$ (Left-To-Right from the edge list). \\
$P_1, P_2, P_3$= These terms are used to weight the contribution of the node in the left, right and path 	from the root to the target node respectively.

\subsubsection{\texorpdfstring{$Level \: l=0 $}{Level l=0}}
 
Number of vertices of a l=0 perfect binary tree is, $v_0=2^0=1$  \\
And number of edges of a l=0 perfect binary tree is, $e_l=(2^l*2)-2 ; e_0=(2^0*2)-2=0$ \\
In a $l=0$ perfect binary tree, there is only one vertex and no edges. Since, there is no edge in this graph, we cannot label this graph. Therefore, it is not possible to label a $l=0$ perfect binary tree. 

\subsubsection{\texorpdfstring{$Level \: l=1 $}{Level l=1}}

In a $l=1$ perfect binary tree, there are $v_1=2^1=2$ vertices and $e_1=(2^1*2)-2=2$ edges. $l=1$ perfect binary tree consists only a root and leaves. According to our approach, a list of prime numbers needs to be built with the same length of edge. Our edge list is, $[2,3]$. Since this perfect binary tree only contains one level, \eqref{eq:1} and \eqref{eq:6} cannot be applied here. For last level, according to \eqref{eq:5} leaves will be, in range $1\geq n \leq 2^l$. \\
\\
For root calculation, according to \eqref{eq:3}:
\begin{align*}
r(1) &= P_1[(2^1 \cdot 2) - 3]^{{th}} + P_2[(2^1 \cdot 2) - 2]^{{th}} \\
r(1) &= P_1(1^{{st}} \text{ value from edge list}) \\
&\quad + P_2(2^{{nd}} \text{ value from edge list}) \\
r(1) &= 2 + 3 = 5
\end{align*}

Apparently, the weights of the vertices are unique Thus, for a  $l=1$ perfect binary tree antimagic labeling is validated by our approach.

\subsubsection{\texorpdfstring{$Level \: l=2 $}{Level l=2}}

For a $l=2$ perfect binary tree, there are $v_2=2^2=4$ vertices and $e_2=(2^2*2)-2=6$ edges. $l=2$ perfect binary tree consists 4 vertices and 6 edges. According to our previously stated approach, a list of prime numbers needs to be created which will have the same length as the edge list. Our edge list is, $[2,3,5,7,11,13]$. As, a $l=2$ level perfect binary tree consists of two levels, equation \eqref{eq:6} cannot be applied here. For last level, according to \eqref{eq:5} leaves will be, in range $1 \geq n \leq 2^l$. \\
\\
For root calculation. According to \eqref{eq:3}:
\begin{align*}
r(2) &= P_1 [(2^2 \cdot 2) - 3]^{th} + P_2 [(2^2 \cdot 2) - 2]^{th} \\
r(2) &= P_1(5)^{th} + P_2(6)^{th} \\
r(2) &= 11 + 13 = 24
\end{align*}
For calculating the value of the second node of the second-to-last level (where $n=2$ and $k=1$), according to \eqref{eq:4}:
\begin{align*}
2(2,1) &= P_1(2 \cdot 2 - 1)^{th} + P_2(2 \cdot 2)^{th} \\
&\quad + P_3[(2^2) + 2]^{th} \\
2(2,1) &= P_1(3)^{rd} + P_2(4)^{th} + P_3[6]^{th} \\
2(2,1) &= 5 + 7 + 13 \\
2(2,1) &= 25
\end{align*}
\begin{figure}[H] 
    \centering
    \includegraphics[height=3.5cm]{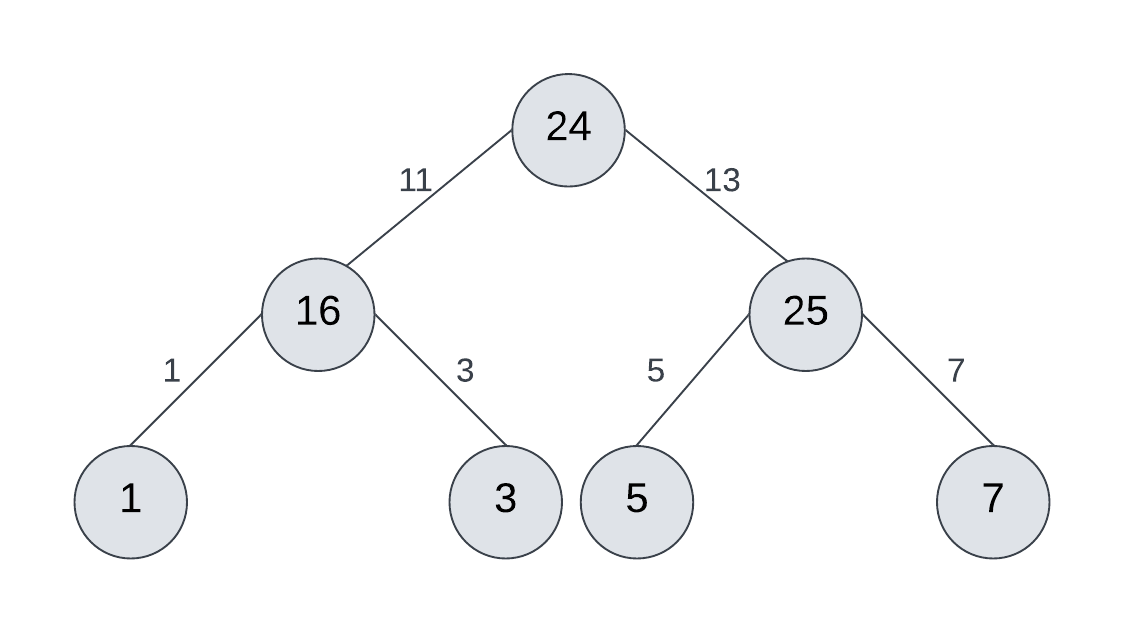}
    \caption{Antimagic labeling for perfect binary tree of level 2}
    \label{fig:3}
\end{figure} 
Apparently, the weights of the vertices are unique Thus, for a  $l=1$ perfect binary tree antimagic labeling is validated by our approach.

\subsubsection{\texorpdfstring{$Level \: l=3 $ and beyond}{Level l=3 and beyond}}

In a l=3 perfect binary tree, there will be a root level, last level, second-to-last level and an internal level. From l=3 we can use all of our equations. There are $v_3=2^3=8$ vertices and $e_3=(2^3*2)-2=14$ edges. Similarly, as $level \: 1 \:$  and $\: level \: 2$, a list of prime numbers needs to be created as the exact same length of edge list. Therefore, our edge list is, $[2,3,5,7,11,13,17,19,23,29,31,37,41,43]$. \\
\\
For root calculation, according to \eqref{eq:3}:
\begin{align*}
r(3) &= P_1[(2^3 \cdot 2) - 3]^{th} + P_2[(2^3 \cdot 2) - 2]^{th} \\
r(3) &= P_1(13)^{th} + P_2(14)^{th} \\
r(3) &= 41 + 43 \\
r(3) &= 84
\end{align*}
For Calculating second-to-last level: $(k=1)$, according to \eqref{eq:4}
\begin{align*}
2(3,1) &= P_1 \left(2 \cdot 2 - 1\right)^{th} + P_2 \left(2 \cdot 2\right)^{th} \\ 
&\quad + P_3 \left(2^3 + 2\right)^{th} \\
2(3,1) &= P_1(3)^{rd} + P_2 (4)^{th} + P_3 (10)^{th} \\
2(3,1) &= 5 + 7 + 29 \\
2(3,1) &= 41
\end{align*}
For Calculating Internal level(where $n=1$ and $k=2$), according to \eqref{eq:6}:
\begin{align*}
1(3,2) = & P_1 \left[(2 \cdot 1-1) + \sum_{i=0}^{0} 2^{3-0}\right]^{th} + \\
& P_2\left[(2\cdot 1) + \sum_{i=0}^{0} 2^{3-0}\right]^{th} + \\
& P_3\left[\sum_{i=0}^{1} 2^{3-0} + 1\right]^{th}
\end{align*}

\begin{align*}
1(3,2) &= P_1(9)^{th} + P_2(10)^{th} + P_3(13)^{th} \\
1(3,2) &= 23 + 29 + 41 = 93
\end{align*}
Likewise, from $level\:3$ and beyond, all the formulas will be applicable. Similarly, a graph of $level\:4$ and $level\:5$ are shown below.

\begin{figure}[H] 
    \centering
    \includegraphics[width=\linewidth]{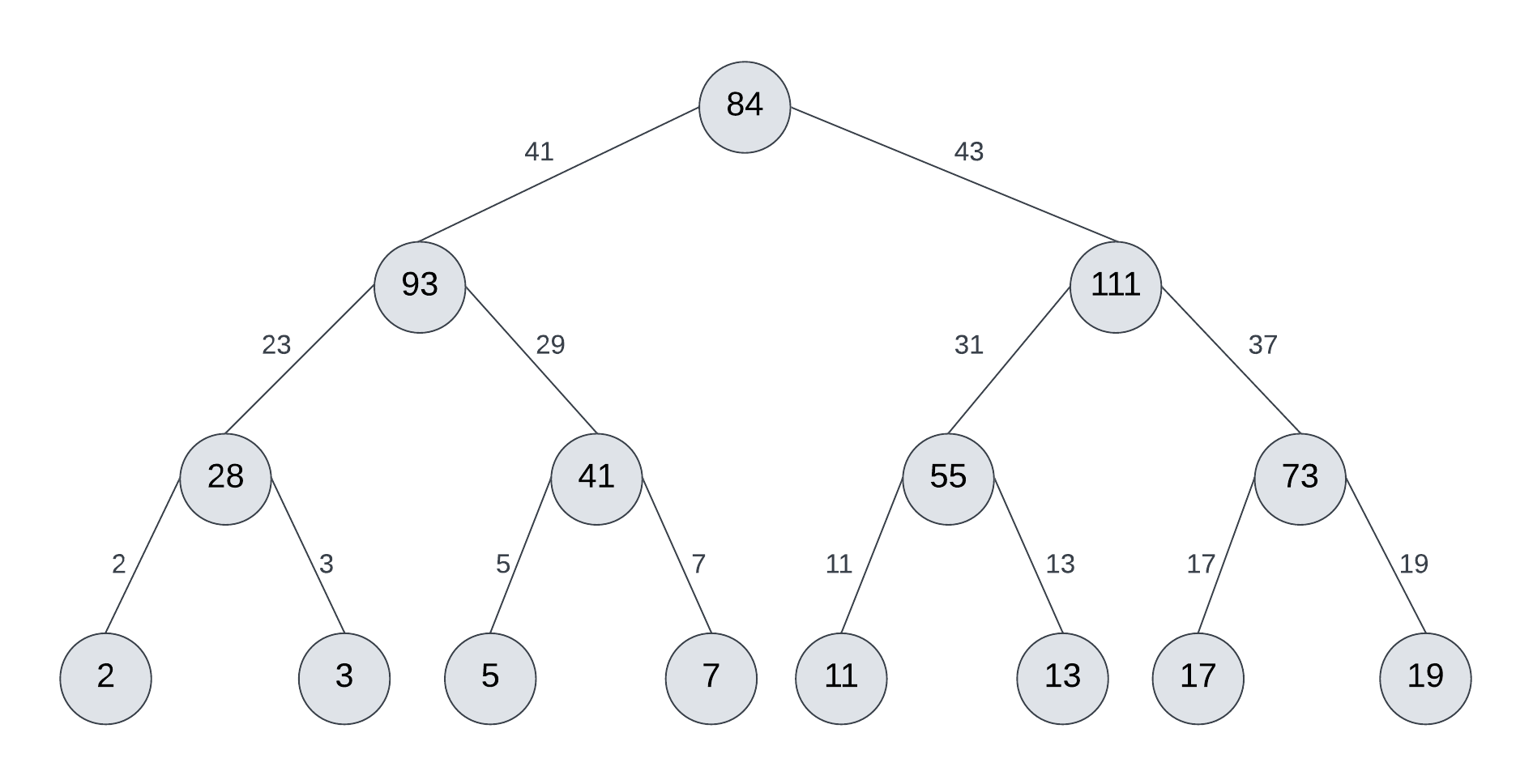}
    \caption{Antimagic labeling for perfect binary tree of level 3}
    \label{fig:4} 
\end{figure} 

\begin{figure}[H] 
    \centering
    \includegraphics[width=\linewidth]{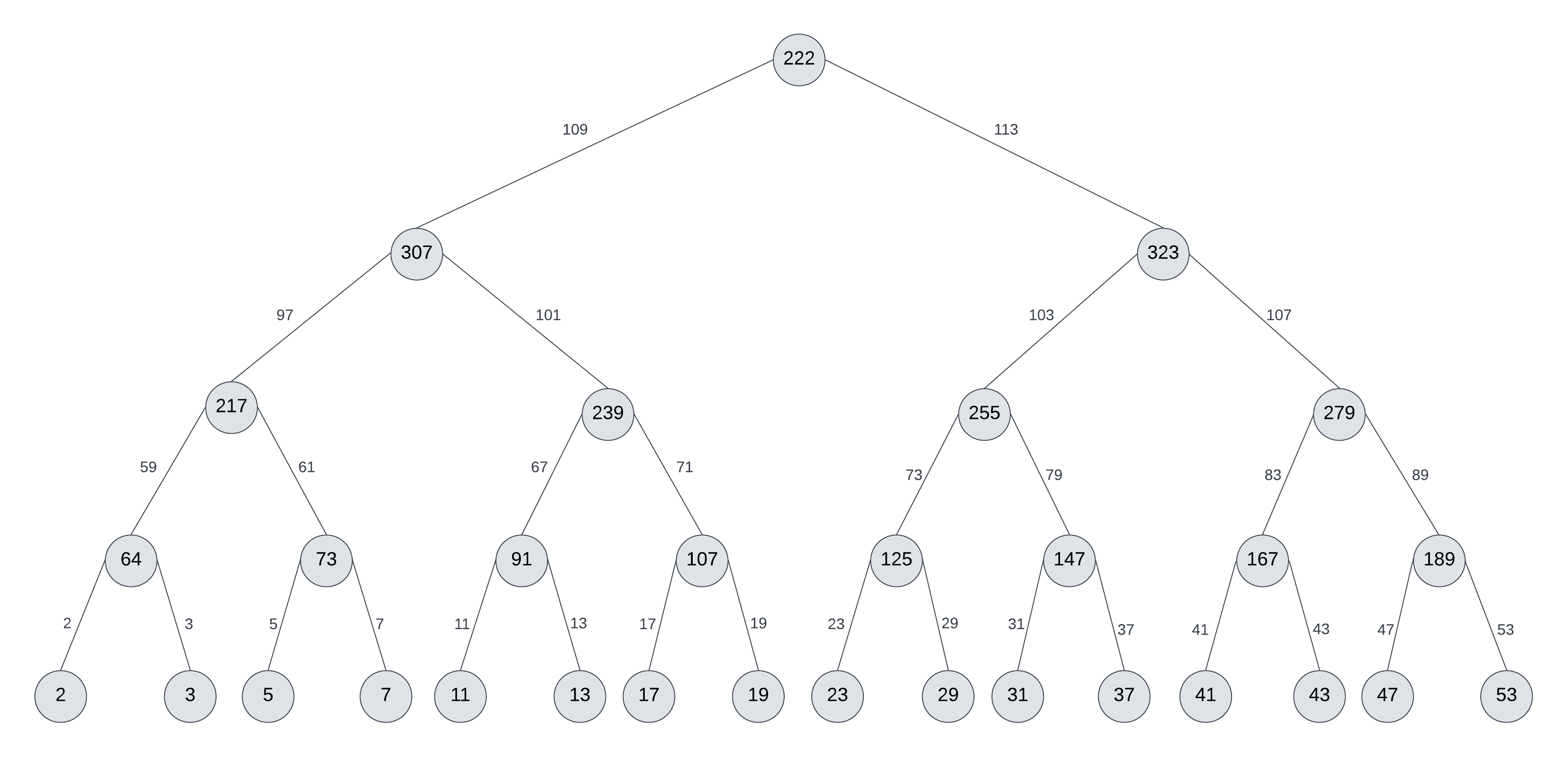}
    \caption{Antimagic labeling for perfect binary tree of level 4}
    \label{fig:5}
\end{figure}

\subsection{Proof:}
\textbf{Theory-1:} A perfect binary tree with its edges labeled using prime numbers, the summation of any three edges of a node is never equal to the sum of another three edges of a different node within the levels between the root and leaves if:
\begin{enumerate}
    \item The edges are ordered in ascending prime values within each node.
    \item At most one prime-numbered edge is shared between the nodes or none.
    \item The sum of the two smaller prime edges in the first node is always less than the sum of the two smaller prime edges in the second node.
    \item The two smaller prime edges within each node are unique.
\end{enumerate} 
\textbf{Proof by Contradiction:} Assume a perfect binary tree with antimagic labeling using prime numbers, and we want to prove that the summation of any three edges of a node is never equal to the sum of another three edges of a different node in the levels between root and leaves.
\begin{figure}[H] 
    \centering
    \includegraphics[width=\linewidth]{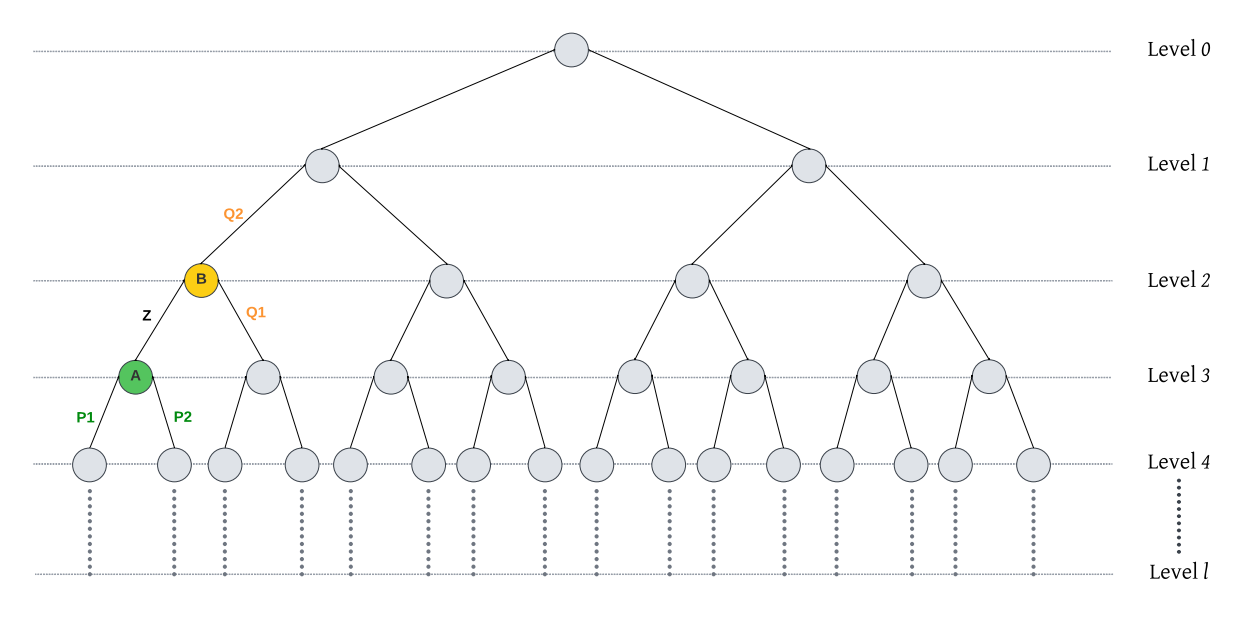}
    \caption{Antimagic labeling for perfect binary tree of $level \ l$}
    \label{fig:theory_1} 
\end{figure} 
With regards to contradiction, that there exist two distinct nodes, node $A$ and node $B$, such that the summation of three edges of node $A$ is equal to the sum of three edges of node $B$: \\
\\
Node $A$'s edges are labeled as $P_1, P_2$, and $Z$, where $P_1 < P_2 < Z$ \\
\\
Node $B$'s edges are labeled as $Q_1, Q_2,$ and $Z$, where $Z < Q_1 < Q_2$ \\
\\
\textit{Equality Assumption:} 

$P_1 + P_2 + Z = Q_1 + Q_2 + Z \ldots \ldots \ldots \ldots (I)$ \\
\\
\textit{Inequality in Edge Labels:}\\ Since the edges are labeled in increasing order, we know that, 

$P_1 < P_2 < Z < Q_1 < Q_2 \ldots \ldots \ldots\ldots \ldots (II)$ \\
\\ Now, let's compare the sums of edges.
\\
\textit{For Node $A$:} 

$P_1 + P_2 + Z < P_2 + P_2 + Z (since P_1 < P_2)$ 

$P_1 + P_2 + Z < 2P_2 + Z \ldots \ldots \ldots \ldots (III)$ \\
\\
\textit{For Node $B$:} 

$Q_1 + Q_2 + Z < Q_2 + Q_2 + Z$ (since $Q_1 < Q_2$) 

$Q_1 + Q_2 + Z < 2Q_2 + Z \ldots \ldots \ldots \ldots (IV)$ \\
\\ Now, comparing $(III)$ and $(IV)$, 

$P_1 + P_2 + Z < 2P_2 + Z < 2Q_2 + Z$ \\

Since $Q_2$ is greater than $P_2$,

$2Q_2 + Z > 2P_2 + Z \ldots \ldots \ldots \ldots (V)$\\
\\
This comparison leads to a contradiction because we initially assumed that $P_1 + P_2 + Z$ equals $Q_1 + Q_2 + Z$, but the above inequalities demonstrate that $P_1 + P_2 + Z$ is strictly less than $Q_1 + Q_2 + Z$.\\
Since our initial assumption led to a contradiction, we conclude that the summation of any three edges of a node is never equal to the sum of another three edges of a different node.


\section{Complete Graph} \label{Complete-Graph}
To demonstrate the feasibility of achieving an antimagic labeling for complete graphs, we have adopted a unique approach likewise perfect binary tree, except it has been proved only by practical implementation. By assigning prime numbers as weights to the tree's edges, we can fulfill all the requirements of antimagic labeling. \\
\\
\textbf{Theory-2:} In a complete graph with n nodes, where each node's value is the sum of all edges connected to that node, and the edges are labeled with prime numbers starting from 2 up to the (n-1)-th prime number, the resulting node values are unique.

\begin{figure}[H]
  \centering
  \includegraphics[width=\linewidth]{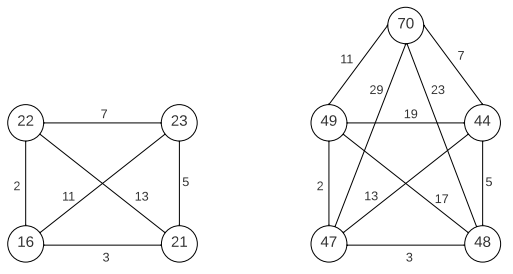}
  \caption{Antimagic labeling of $K_4$ (left) and $K_5$ (right) complete graph}
  \label{fig:7}
\end{figure}

\subsection{Pseudocode:}
\begin{algorithm} [H]
\caption{Generate Complete Graph With Antimagic Labeling}
\begin{algorithmic}[1]
\State $G \gets \text{create\_complete\_graph}(n)$
\State $prime\_edges \gets \text{get\_prime\_edges}(n)$
\For{$i$ from $1$ to $n$}
    \For{$j$ from $i+1$ to $n$}
        \State $\text{add\_edge}(G, i, j)$
        \State $\text{label\_edge}(G, i, j, prime\_edges[i-1])$
    \EndFor
\EndFor

\For{$i$ from $1$ to $n$}
    \State $node\_value \gets 0$
    \For{each edge $(u,v)$ in $\text{edges\_of}(G)$}
        \If{$u$ is $i$ or $v$ is $i$}
            \State $label \gets \text{get\_edge\_label}(G, u, v)$
            \State $node\_value \gets node\_value + label$
        \EndIf
    \EndFor
\EndFor
\end{algorithmic}
\end{algorithm}

\subsection{Proof:}
\textbf{Proof by Contradiction: }\\
Let's assume the opposite, that there exist two distinct nodes, Node A and Node B, with the same node value in such a complete graph. We denote the set of edges connected to Node A as ${2, 3, 5, ..., p}$, where $p$ is the $(n-1)^{th}$ prime number. Similarly, the set of edges connected to Node B is denoted as ${2, 3, 5, ..., q}$, where q is also the $(n-1)^{th}$ prime number. \\
\\
Since $Node \: A$ and $Node \: B$ have the same node value, we can express this as: \\
$Node \: A's$ Value = $Node \: B's$ Value \\
Mathematically, this can be written as below: \\
$\displaystyle\sum\limits_{i=2}^p i = \displaystyle\sum\limits_{i=2}^q i$ \\
\\
Where p and q are both equal to the (n-1)th prime number. \\
Now, let's consider the sum of all edges connected to Node A minus the sum of all edges connected to Node B: \\
$\displaystyle\sum\limits_{i=2}^p i - \displaystyle\sum\limits_{i=2}^q i = 0$\\
\\
Mathematically, this can be written as: \\
$\displaystyle\sum\limits_{i=2}^p (i - j) = 0$ \\
\\
However, since each edge is labeled with a unique prime number starting from 2, we have: \\
$(i-j)\neq 0$ \\
\\
So, for each $i\neq 0$, we have: \\
$\displaystyle\sum\limits_{i=2}^p i - \displaystyle\sum\limits_{i=2}^q i \neq 0$ \\
\\
This means that there must be at least one term (i-j) in the sum that is not equal to 0.\\
\\
This contradicts our initial assumption that the node values of Node A and Node B are equal. Therefore, our assumption is false, and we conclude that in a complete graph with n nodes, where each node's value is the sum of all edges connected to that node, and the edges are labeled with prime numbers starting from $2$ up to the $(n-1)^{th}$ prime number, the resulting node values are indeed unique.

\section{Antimagic Labeling of Other Graphs} \label{Other-Graphs}
\begin{figure}[H]
  \centering
  \includegraphics[width=4cm]{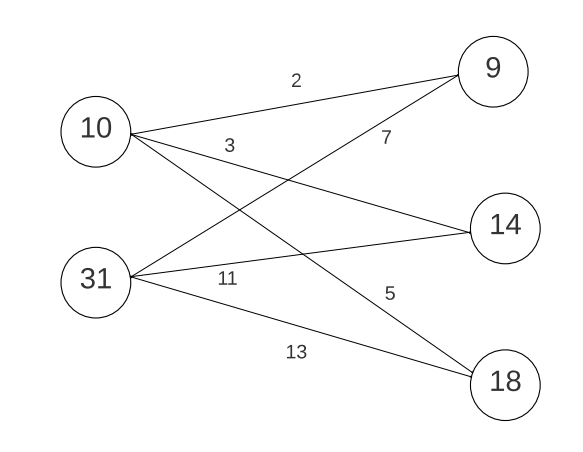}
  \caption{$K_{2,3}$ cubic bipartite graph (special case)}
  \label{fig:12}
\end{figure}

While doing this research a noteworthy issue has come to our attention. We found that sometimes multiple nodes end up with the same values when we try to label the edges arbitrarily, which doesn't meet the antimagic labeling criteria. This issue is most pronounced when nodes in a graph are connected by three edges.We propose a solution that avoids the issue of overlapping values across graphs. If edges of the graphs are labeled in an orderly fashion, the conditions of antimagic labeling are not violated. It can be ascending or descending in order. With this approach, we ensure that no two nodes have the same value, satisfying the conditions for antimagic labeling. In case of non-regular bipartite graph, if it is labeled using prime numbers, it will always be antimagic labeling and it does not require to follow any order while giving weights to the edges (see figure:\ref{fig:12}). Another exceptional case arises in ladder graph, when the number of nodes less than or equal four $(n \leq 4)$, it will always follow the conditions of antimagic labeling (see figure:\ref{fig:12}). However, if the number of nodes of a ladder graph are less than or equal to two $(n \leq 4)$, maintain the conditions of antimagic labeling will not be feasible. Our research not only identifies all these issues and offers a practical solution, especially when nodes have three edges connecting them in the graph.
Few examples of various graphs with their problems and solutions are graphically illustrated below. For each graph figure at the left represents with its problem and at the right it is shown with solutions. \\
\\ \textbf{Graphs With Problems(left) and Solutions(right):}

\begin{figure}[H]
  \centering
  \includegraphics[width=\linewidth]{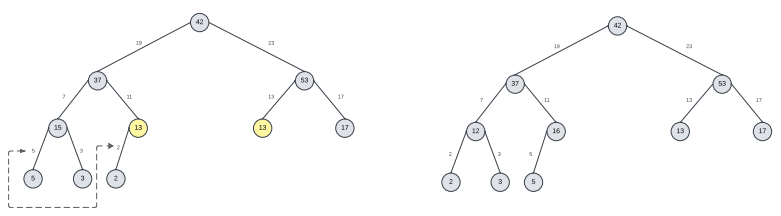}
  \caption{Complete binary trees of level 3}
  \label{fig:8}
\end{figure}

\begin{figure}[H]
  \centering
  \includegraphics[width=\linewidth]{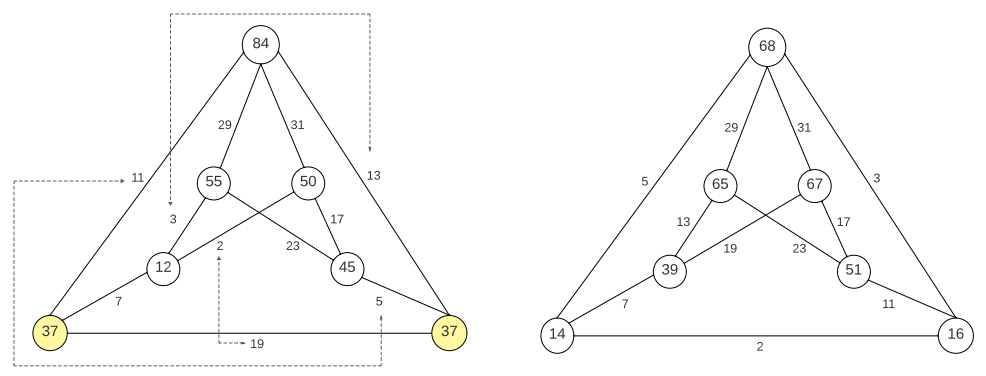}
  \caption{$W_7$ pyramid graphs.}
  \label{fig:9}
\end{figure}

\begin{figure}[H]
  \centering
  \includegraphics[width=\linewidth]{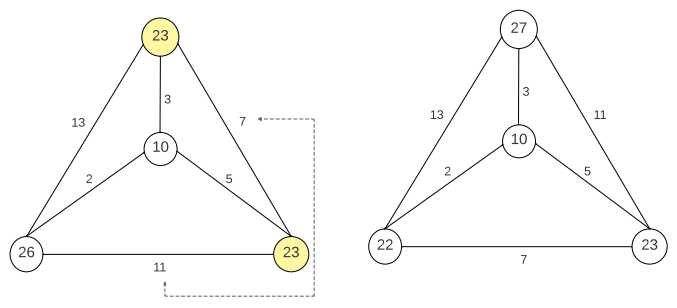}
  \caption{$K_4$ Cubic graphs}
  \label{fig:10}
\end{figure}

\begin{figure}[H]
  \centering
  \includegraphics[width=\linewidth]{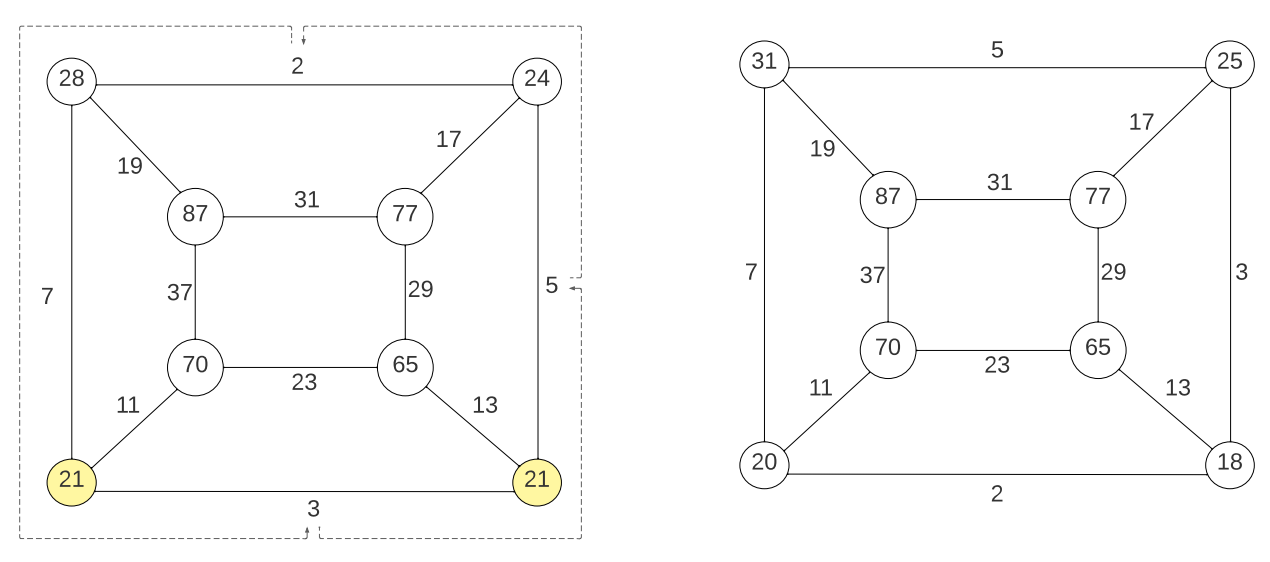}
  \caption{$Q_3$ cubic graphs.}
  \label{fig:11}
\end{figure}

\begin{figure}[H]
  \centering
  \includegraphics[width=\linewidth]{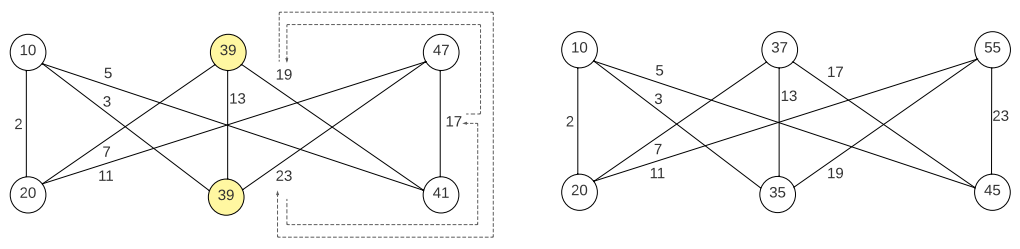}
  \caption{$K_{3,3}$ complete bipartite graphs.}
  \label{fig:13}
\end{figure}

\begin{figure}[H]
  \centering
  \includegraphics[width=\linewidth]{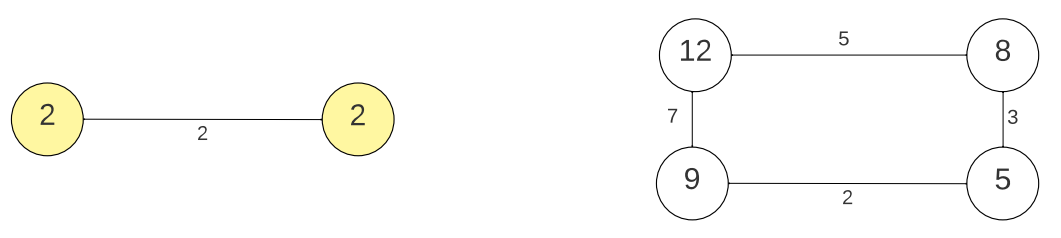}
  \caption{${}_1 P_2$ (left) and ${}_2 P_2$ (right) ladder graphs (special cases).}
  \label{fig:14}
\end{figure}

\begin{figure}[H]
  \centering
  \includegraphics[width=\linewidth]{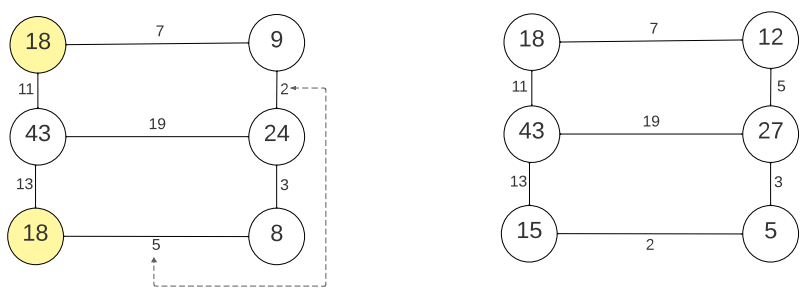}
  \caption{${}_3 P_2$ ladder graphs.}
  \label{fig:15}
\end{figure}

\section{Results and Analysis} \label{results}
In this paper, we have mathematically verified that antimagic labeling is possible for any perfect binary tree using prime numbers. From the table, we can see that the values of the roots are absolutely unique. Thus, we can say that, if a perfect binary tree of $level \: l$ is labeled from $1$ to $e_l$ progressively from left-to-right and bottom-to-top using prime numbers, antimagic labeling is possible for any perfect binary tree. In the table, weights of nodes’, roots and total number of nodes of specific level are provided. All the weights of any nodes of any level are completely distinctive. Additionally, the number of edges and nodes increase exponentially with the number of levels. Values of only 24 levels are provided because of compact area. Beyond the empirical results, this research significantly contributes to the broader comprehension of antimagic labelling within the context of perfect binary trees. Furthermore, it expands the existing body of knowledge in the expansive field of graph labelling, accentuating its relevance and significance within the realm of mathematics and computer science.

In the table: 1, first three leftmost vertices of $level \: (l-1)$, first 2 leftmost vertices of $level \: (l-2)$ and the leftmost vertices of $l=3$ are given along with the values of root nodes and total number of nodes.

\begin{table*}
\centering
\caption{Weights of certain vertices of particular levels of perfect binary tree.}
\vspace{10pt} 
\label{tab:my-table} 
\resizebox{1\textwidth}{!}{%
\begin{tabular}{|l|l|l|l|l|l|l|l|l|l|}
\hline 
$\textbf{Level,\:l}$ & $\textbf{w1,\:l-1}$ & $\textbf{w2,\:l-1}$ & $\textbf{w3,\:l-1}$ & $\textbf{w1,\:l-2}$ & $\textbf{w2,\:l-2}$ & $\textbf{w3,\:l-2}$ & $\textbf{w1,\:l-3}$ & $\textbf{Root \: value}$ & $\textbf{No. \: of \: Nodes}$ \\ \hline
0 & - & - & - & - & - & - & - & 2 & 1 \\ \hline
1 &  &  & - & - & - & - & - & 5 & 3 \\ \hline
2 & 16 & 25 & - & - & - & - & - & 24 & 7 \\ \hline
3 & 28 & 41 & 55 & 93 & 111 & - & - & 84 & 15 \\ \hline
4 & 64 & 73 & 91 & 217 & 239 & 255 & 307 & 222 & 31 \\ \hline
5 & 142 & 151 & 173 & 503 & 529 & 553 & 725 & 576 & 63 \\ \hline
6 & 318 & 329 & 355 & 1139 & 1189 & 1219 & 1647 & 1392 & 127 \\ \hline
7 & 732 & 745 & 763 & 2631 & 2663 & 2695 & 3779 & 3216 & 255 \\ \hline
8 & 1626 & 1639 & 1661 & 5907 & 5957 & 6001 & 8491 & 7280 & 511 \\ \hline
9 & 3678 & 3689 & 3715 & 13201 & 13245 & 13271 & 18685 & 16240 & 1023 \\ \hline
10 & 8172 & 8183 & 8203 & 29249 & 29287 & 29347 & 41177 & 35676 & 2047 \\ \hline
11 & 17886 & 17903 & 17927 & 63955 & 64013 & 64043 & 89871 & 77712 & 4095 \\ \hline
12 & 38896 & 38915 & 38941 & 138755 & 138839 & 138863 & 194431 & 267970 & 8191 \\ \hline
13 & 84052 & 84065 & 84083 & 299643 & 299681 & 299737 & 418823 & 360964 & 16383 \\ \hline
14 & 180516 & 180545 & 180563 & 642763 & 642817 & 642857 & 896883 & 772098 & 32767 \\ \hline
15 & 386122 & 386131 & 386153 & 1372939 & 1372987 & 1373043 & 1911961 & 1643124 & 65535 \\ \hline
16 & 821652 & 821663 & 821687 & 2919047 & 2919091 & 2919267 & 4058611 & 3485014 & 131071 \\ \hline
17 & 1742544 & 1742575 & 1742603 & 6184563 & 6184641 & 6184689 & 8586745 & 7362108 & 262143 \\ \hline
18 & 3681154 & 3681163 & 3681215 & 13056451 & 13056553 & 13056643 & 18107685 & 15508020 & 524287 \\ \hline
19 & 7754086 & 7754125 & 7754143 & 27481693 & 27481769 & 27481885 & 38069135 & 32580032 & 1048575 \\ \hline
20 & 16290078 & 16290101 & 16290143 & 57697399 & 57697481 & 57697523 & 79843061 & 68272008 & 2097151 \\ \hline
21 & 34136064 & 34136089 & 34136107 & 120840189 & 120840259 & 120840355 & 167071827 & 142757070 & 4194303 \\ \hline
22 & 71378608 & 71378633 & 71378665 & 252538565 & 252538645 & 252538709 & 348849659 & 297896236 & 8388607 \\ \hline
23 & 148948146 & 148948169 & 148948195 & 526748179 & 526748251 & 526748303 & 727091047 & 620496456 & 16777215 \\ \hline
24 & 310248256 & 310248355 & 310248371 & 1096697093 & 1096697219 & 1096697297 & 1512761729 & 1290310356 & 33554431 \\ \hline
\end{tabular}%
}
\end{table*}

\section{Applications of Antimagic Labeling for Graphs} \label{Applications} 

\textbf{Graph Theory:} Prime-number-based antimagic labeling allows for the study of unique properties and advantages, particularly in the study of graph structures. For instance, it opens the window to study of graph isomorphism, a fundamental problem in graph theory. Consider two non-isomorphic graphs, $G_1$ and $G_2$. By applying prime-number-based antimagic labeling to them, $G_1$ receives labels like $2, 3, 5$, and $7$, while $G_2$ is labeled differently, say, $11, 13, 17$, and $19$. The distinctiveness of the prime labels signifies that $G_1$ and $G_2$ are structurally different, providing valuable information in graph isomorphism testing. This technique may also be helpful for the study of Hamiltonian cycles and Eulerian circuits. \\
\\ \textbf{Binary Search Trees: } Antimagic labeling using prime numbers optimizes the efficiency of binary search trees (BSTs), reducing search complexity and improving data retrieval efficiency. For example, in a perfect binary search tree labeled with prime numbers, the labels may represent the magnitude of elements stored in the tree, leading to efficient data retrieval, particularly in database systems.\\
\\ \textbf{Cryptography:} Prime numbers have been a cornerstone of cryptographic methods for their unpredictability and difficulty to factor. Antimagic labeling using primes can provide an extra layer of security by creating prime-labeled structures that are challenging for adversaries to decipher. The inherent properties of prime numbers can make encrypted data even more resistant to attacks, providing a level of security that other approaches might not achieve.\\
\\ \textbf{Parallel Computing:} Today there is a sharp increase in the demand for faster and more efficient processing of complex tasks. Antimagic labeling with prime numbers can assist in load balancing and task distribution in parallel and distributed computing environments. Prime-number-based labels can signify the computational load of individual nodes in a distributed computing system, facilitating balanced load distribution.
Algorithm Design: Antimagic labeling using prime numbers can influence algorithm design by offering a unique labeling scheme with interesting arithmetic properties. Algorithms for optimization, sorting, or search problems can incorporate prime-number-based labels for improved efficiency and robustness.\\
\\ \textbf{Artificial Intelligence:} In AI, prime-number-based antimagic labeling can be applied to decision tree-based machine learning algorithms. Prime-number-based labels on nodes in a decision tree can represent the significance of different features in a dataset, allowing for more efficient and accurate classification in AI applications. Prime-number-based labels can be integrated into the training data for supervised learning tasks. Models can learn to utilize these labels for better feature discrimination and improved classification or regression outcomes.\\
\\ \textbf{Graph Neural Networks (GNNs):} In the context of GNNs, prime-number-based antimagic labeling using prime numbers can be used to enhance graph representation learning. GNNs can leverage prime-number-based node labels to improve node embeddings and community detection in complex networks, leading to better graph analysis and prediction.\\

\section{Limitations} \label{Limitations}
As the number of levels increase the number of nodes and edges also increase exponentially. Google Colab version 3 is used to implement in this research. Therefore, after executing 24 levels, google collab restricts us from using it. Because of having limited memory, the execution could not be carried out further. In simpler terms, the main limitation we faced was that our computations couldn't run indefinitely on the free version of Google Colab. \\
The next limitation is, we could not build a universal formula for any level, rather, we have built 4 distinguish formulas. In the context of cubic graphs, it is observed that the arbitrary assignment of labels to the edges may lead to a violation of the conditions for antimagic labeling, resulting in the inadvertent occurrence of edge-label collisions between two distinct nodes. For instance, if we label the edges of a node $(11, 5, 2)$ and $(13, 3, 2)$ for the other node, the values of two nodes become the same.

\section{Conclusions} \label{Conclusions}
This exploration of antimagic labeling utilizing prime numbers represents a promising approach to addressing graph labeling problems. The use of prime numbers offers a versatile and potentially efficient method for assigning labels to graphs. We have shown antimagic labeling of any perfect binary tree and complete graphs using prime without any exception. We also demonstrated antimagic labeling for bipartite graphs, cubic graphs, ladder graphs, binary trees, complete binary trees and pyramid graphs with few exceptions. However, we have shown that those problems can be solved by following a sequential order. This research also highlights the potential of antimagic labeling for broader applications in various fields, such as artificial intelligence and cryptography, among others. As the study of antimagic labeling continues to evolve, it opens new areas for solving complex problems in diverse mathematical and computational domains.

\bibliographystyle{unsrt}  

\bibliography{references.bib}





\end{multicols}
\end{document}